\documentclass[hidelinks]{ifacconf}

\usepackage{filecontents}

\makeatletter
\let\old@ssect\@ssect 
\makeatother

\usepackage{graphicx}      
\usepackage{natbib}        
\usepackage{amsmath}
\usepackage{amssymb}
\usepackage{mathtools}
\usepackage{bm}
\usepackage{hyperref}
\usepackage{booktabs}
\usepackage{xcolor}
\usepackage{doi}

\def\equationautorefname~#1\null{%
  (#1)\null
}

\DeclareMathOperator\diag{diag}

\newcommand{\R}{\mathbb{R}}

\newcommand{\sys}{\bm{\Sigma}}          

\renewcommand{\S}{S}

\newcommand{\sysc}{\check{\sys}} 

\newcommand{\Ac}{\check{A}}
\newcommand{\Bc}{\check{B}}
\newcommand{\Cc}{\check{C}}

\newcommand{\Pc}{\check{P}}
\newcommand{\Qc}{\check{Q}}
\newcommand{\Pic}{\check{\Pi}}
\newcommand{\Xic}{\check{\Xi}}

\newcommand{\sysb}{\tilde{\sys}} 

\newcommand{\Ab}{\tilde{A}}
\newcommand{\Bb}{\tilde{B}}
\newcommand{\Cb}{\tilde{C}}
\newcommand{\Db}{\tilde{D}}

\newcommand{\Xib}{\tilde{\Xi}}

\newcommand{\sysr}{\hat{\bm{\Sigma}}}
\newcommand{\Gr}{\hat{G}}

\newcommand{\Ar}{\hat{A}}
\newcommand{\Br}{\hat{B}}
\newcommand{\Cr}{\hat{C}}
\newcommand{\Dr}{\hat{D}}

\newcommand{\Xir}{\hat{\Xi}}

\makeatletter
\def\@ssect#1#2#3#4#5#6{%
  \NR@gettitle{#6}
  \old@ssect{#1}{#2}{#3}{#4}{#5}{#6}
}
\makeatother

\begin{document}
\begin{frontmatter}

\title{Passivity-Preserving, Balancing-Based Model Reduction for Interconnected Systems \thanksref{footnoteinfo}} 

\thanks[footnoteinfo]{This project was supported by ASML and the TKI program of the Dutch Government.}

\author[First]{Luuk Poort} 
\author[Second]{Bart Besselink} 
\author[First]{Rob H. B. Fey}
\author[First]{Nathan van de Wouw}

\address[First]{Eindhoven University of Technology, 
   Eindhoven, The Netherlands (e-mail: l.poort@tue.nl, r.h.b.fey@tue.nl, n.v.d.wouw@tue.nl).}
\address[Second]{Bernoulli Institute for Mathematics, Computer Science and Artificial Intelligence, University of Groningen, Groningen, The Netherlands (e-mail: b.besselink@rug.nl).}

\begin{abstract}                
    This paper proposes a balancing-based model reduction approach for an interconnection of passive dynamic subsystems. This approach preserves the passivity and stability of both the subsystems and the interconnected system. Hereto, one Linear Matrix Inequality (LMI) per subsystem and a single Lyapunov equation for the entire interconnected system needs to be solved, the latter of which warrants the relevance of the reduction of the subsystems for the accurate reduction of the interconnected system, while preserving the modularity of the reduction approach. In a numerical example from structural dynamics, the presented approach displays superior accuracy with respect to an approach in which the individual subsystems are reduced independently.\vspace{-2mm}
\end{abstract}


\end{frontmatter}

\section{Introduction}
Highly complex models, e.g., RLC networks, integrated circuits or structural dynamics systems, can often be regarded as an interconnection of several subsystems. To meet accuracy requirements, each subsystem is typically described by a high-order model. The large order of the resulting interconnected system prevents the use of computationally costly techniques for controller synthesis or observer design. Therefore, it is often required to first approximate the high-order model with a surrogate model of lower order.

The search for proper approximate models is the main goal of the Model Reduction (MR) field. One of the most popular MR methods is Balanced Truncation (BT), as originally presented by \cite{Moore1981PrincipalReduction}. It has gained this popularity due to its simplicity, accuracy, stability guarantee and availability of an a priori error bound \citep{Antoulas2005ApproximationSystems}. Unfortunately, other important properties of the original system, in particular passivity and the internal structure of a system, are not necessarily preserved.

The topic of preserving passivity of a system with BT has received considerable attention since the first works of \cite{Jonckheere1983ADesign} and \cite{Desai1984AReduction}. In subsequent research, \cite{Unneland2007NewTruncation} presented sufficient conditions to preserve passivity, making the reduction more efficient and flexible. Further extensions by \cite{Zulfiqar2017ATechnique} and \cite{Imran2018FrequencyTechnique} also allow the user to shape the approximate model by frequency weighting.

To preserve the internal interconnection structure of interconnected systems, model reduction is usually performed on the individual subsystems. This approach does not take the dynamics of the interconnected system into account when approximation subsystems, such that the interconnected system might actually be approximated poorly \citep{Sandberg2009}. Although frequency-weighted reduction of the subsystems can improve the approximation of the interconnected system, this presents the issue of designing appropriate weighting. \cite{Vandendorpe2008ModelSystems} propose a more elegant solution based on BT, reducing the subsystems based on the dynamics of the interconnected system. This effectively reduces the interconnected system, while retaining its internal structure. 



Although both preservation of passivity and preservation of internal structure with BT are adequately researched seperately, their combination has received limited attention. While \cite{Cheng2019BalancedSystems} present a BT method for an interconnection of identical systems to retain structure and passivity, there is no BT method to retain this structure and passivity for general interconnected subsystems.

In this paper, we combine the passivity preservation of \cite{Unneland2007NewTruncation} with the preservation of internal structure, as done by \cite{Vandendorpe2008ModelSystems}, into a new BT method for interconnected systems. More explicitly, the proposed method solves one Linear Matrix Inequality (LMI) per subsystem and a single Lyapunov equation for the interconnected system to retain accuracy of the reduced interconnected system, while guaranteeing the passivity of both the reduced subsystems and the interconnected system. 

The paper is organized as follows. In \autoref{sec:prob_def}, the problem statement will be defined in detail. The general concept of balanced truncation is subsequently treated in \autoref{sec:bal_trunc}. In \autoref{sec:interc_systems}, the proposed method of passive interconnected balanced truncation is introduced. Then, this new method is illustrated by means of a numerical example in \autoref{sec:num_example}. Finally, conclusions on its applicability are drawn in \autoref{sec:conclusion}.

\emph{Notation:}
The field of all real numbers is denoted by $\R$. $\R^n$ and $\R^{n\times p}$ indicate an $n$-dimensional vector and $n$ by $p$ real matrix, respectively. $\R_{> 0}$ denotes the set of positive real numbers and $O$ and $I$ denote the zero and identity matrices, respectively. If a matrix $A$ is positive definite, we indicate this as $A\succ 0$. Similarly, negative, semi-positive and semi-negative definiteness are indicated as $A\prec 0$, $A\succeq 0$ and $A\preceq 0$. $A \succ B$ implies $A-B \succ 0$.

\vspace{-1mm}
\section{Problem statement}\label{sec:prob_def}\vspace{-2mm}

Consider the square, minimal state-space model
\begin{equation}\label{eq:default_SS}
    \sys:
    \begin{cases}
        \dot{x} = A x + B u,\\
        y = C x + Du,
    \end{cases}
\end{equation}
also denoted as $\sys\coloneqq(A,B,C,D)$ with state $x\in\R^n$, input $u\in\R^p$ and output $y\in\R^p$. In this paper, we will often consider $\sys$ to be passive.

\begin{defn}\label{def:passsivity} 
\citep{Willems1972DissipativeTheory}
    A minimal system $\sys$ as in \autoref{eq:default_SS} is \emph{passive} if there exists a positive \emph{storage} function $H(x):\R^n\rightarrow\R_{>0}$, such that for all $t_0 \leq t_1$, and along all trajectories of \autoref{eq:default_SS}, we have
    \begin{equation}\label{eq:pass_ineq}
        H(x(t_0)) + \int_{t_0}^{t_1} y(\tau)^\top u(\tau) d\tau \geq H(x(t_1)).
    \end{equation}
\end{defn}

\begin{lem}\label{lem:pos_real}
\emph{Positive Real Lemma} \citep{Antoulas2005ApproximationSystems, Willems1972DissipativeRates}:
    The square, minimal system $\sys$ of \autoref{eq:default_SS} is passive if and only if there exists a matrix $\Xi = \Xi^\top\succ0$ such that
    \begin{equation}\label{eq:pass_LMI}
        \left[\begin{array}{cc}
            A^{\top} \Xi+\Xi A & \Xi B-C^\top \\
            B^\top \Xi-C & -(D+D^\top)
        \end{array}\right] \preceq 0.
    \end{equation}
\end{lem}

\begin{rem}
    Note that $\Xi$ in Lemma \ref{lem:pos_real} is positive definite because $\sys$ is observable, as shown in \cite[Lemma~1]{Willems1972DissipativeRates}. Any system that is passive according to Lemma \ref{lem:pos_real} is therefore also Lyapunov stable.
\end{rem}

Now consider $k$ passive, minimal subsystems $\sys_1,\dots, \sys_k$, where $\sys_j \coloneqq (A_j,B_j,C_j,D_j)$, with states $x_j\in\R^{n_j}$, inputs $v_j\in\R^{p_j}$ and outputs $z_j\in\R^{p_j}$, $j\in\{1,\dots,k\}$.

The system $\sys_b \coloneqq (A_b,B_b,C_b,D_b) = \diag(\sys_1,\dots,\sys_k)$ is a parallel composition of subsystems $\sys_1,\dots,\sys_k$, such that
\begin{equation}
    \begin{aligned}
        A_b &= \diag(A_1,\dots,A_k),& B_b &= \diag(B_1,\dots,B_k),\\
        C_b &= \diag(C_1,\dots,C_k),& D_b &= \diag(D_1,\dots,D_k),
    \end{aligned}\\
\end{equation}
with system states $x_b \in \R^{n_b}$, inputs $v_b \in \R^{p_b}$ and outputs $z_b \in \R^{p_b}$ as
\begin{equation}\label{eq:state&IO_partitions}
    \begin{gathered}
        x_b = \begin{bmatrix}x_1\\\vdots\\x_k \end{bmatrix},\quad
        v_b = \begin{bmatrix}v_1\\\vdots\\v_k \end{bmatrix},\quad
        z_b = \begin{bmatrix}z_1\\\vdots\\z_k \end{bmatrix}.
    \end{gathered}
\end{equation}
\begin{rem}\label{rem:sysb_passivity}
    $\sys_b$ is also passive, because parallel interconnections of passive subsystems are passive, as shown by, e.g., \cite{Bao2007ProcessApproach}, Theorem 2.33. 
\end{rem}

\begin{figure}
    \centering
    \includegraphics[width = 0.75\linewidth]{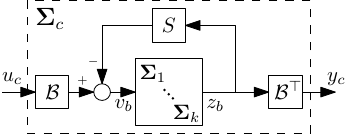}
    \caption{Schematic representation of the interconnected system}
    \label{fig:cpld_diag}
\end{figure}

Observe a negative feedback interconnection of $\sys_b$ with the positive semi-definite interconnection matrix $S\in \R^{p_b\times p_b}$ such that $I+S D_b$ and $I+D_b S$ are nonsingular, as shown in \autoref{fig:cpld_diag}. This interconnection is pre- and post-multiplied with the matrices $\mathcal{B}\in\R^{p_b\times p_c}$ and $\mathcal{B}^\top$, respectively, such that 
\begin{equation}
    v_b = -Sz_b + \mathcal{B}u_c, \qquad y_c = \mathcal{B}^\top z_b,
\end{equation}
where $u_c, y_c \in \R^{p_c}$ are the inputs and outputs of the interconnected system $\sys_c$, with states $x_c \in \R^{n_c}, n_c = n_b$.

Hence, the interconnected system is given as
\begin{equation}\label{eq:cpld_sys}
    \sys_c :
    \begin{cases}
        \dot{x}_c= A_c x_c + B_c u_c,\\
        y_c\hspace{0.75pt} = C_c x_c + D_cu_c,
    \end{cases} 
\end{equation}
with system matrices
\begin{equation}
    \begin{aligned}\label{eq:cpld_sys_matrices}
        A_c &= A_{b}-B_{b} \mathcal{D}_2 S C_{b}, &
        B_c &= B_{b} \mathcal{D}_{2} \mathcal{B},\\ 
        C_c &= \mathcal{B}^\top \mathcal{D}_{1} C_b , &
        D_c &= \mathcal{B}^\top D_{b} \mathcal{D}_{2} \mathcal{B}, \\
        \mathcal{D}_1 &= (I+D_b S)^{-1}, &
        \mathcal{D}_2 &= (I+S D_b)^{-1}.
    \end{aligned}
\end{equation}

It is easy to show that the interconnected system $\sys_c$ is itself passive.
\begin{thm}\label{th:interconnection_passivity}
    Given the passive system $\sys_b$ with $p_b$ inputs and outputs, positive semi-definite matrix $S \in \R^{p_b\times p_b}$ and matrix $\mathcal{B}\in \R^{p_b\times p_c}$, the interconnected system $\sys_c$ with $p_c$ inputs $u_c$ and outputs $y_c$, as given in \autoref{eq:cpld_sys}, is passive.
\end{thm}
\begin{pf}
    Since $\sys_b$ is passive (Remark \ref{rem:sysb_passivity}), the inequality of \autoref{eq:pass_ineq} holds for the inputs $v_b$ and outputs $z_b$. Substitution of the coupling equations $v_b = -Sz_b + \mathcal{B}u_c$ and $y_c = \mathcal{B}^\top z_b$ gives
    \vspace{-2mm}
    \begin{equation}
        \begin{split}
            H(x_b(t_0)) + \int_{t_0}^{t_1} y_c(\tau)^\top u_c(\tau) d\tau  \hspace{25mm}\\
            \hspace{25mm}- \int_{t_0}^{t_1} z_b(\tau)^\top S z_b(\tau) d\tau \geq H(x_b(t_1)),
        \end{split}
    \end{equation}
    such that $\sys_c$ is also clearly passive for $S\succeq 0$ with inputs $u_c$, outputs $y_c$ and storage function $H(x_c) = H(x_b)$.
\hfill $\square$ \end{pf}

Note that it is also possible to treat $S$ as a passive system of only direct feedthrough, according to Lemma \ref{lem:pos_real}. The result that a negative feedback interconnection of two passive systems is also passive, is actually well-known in literature (e.g., \cite{Willems1972DissipativeTheory} and \cite{Zhu2017OnIndices}). 


In this paper, we will approximate the subsystem models $\sys_j$ by reduced-order models $\sysr_j$ for all $j\in \{1,\dots,k\}$, which makes for a modular reduction approach. 
Analogous to the unreduced case, the parallel composition $\sysr_b = \diag(\sysr_1,\dots,\sysr_k)$ is interconnected with $S$ and $\mathcal{B}$, using the equations of \autoref{eq:cpld_sys_matrices}, to form the reduced interconnected model $\sysr_c$.

The goal of the reduction is to reduce passive subsystems $\sys_j$ to $\sysr_j$ for all $j\in \{1,\dots,k\}$, such that
\begin{enumerate}
    \item the subsystems $\sysr_j$ are passive and stable.
    \item the frequency response function of $\sysr_c$ accurately approximates the frequency response function of $\sys_c$.
    \item $\sysr_c$ is passive and stable.
\end{enumerate}



\section{Balanced truncation: Background}\label{sec:bal_trunc}
Balanced truncation (BT) is a model reduction procedure to generate lower-order surrogate models.
In this section, we give preliminary information about BT in its application to a single system of type \autoref{eq:default_SS}. Specifically, we will introduce Lyapunov, positive real and mixed Gramian balancing as the basis for proposal of the new method in \autoref{sec:interc_systems}. 

The procedure of BT methods is based on two steps: first the coordinates are transformed to a balanced realisation, where states are sorted by their `importance', followed by a truncation of the least important states. This `importance' can be specified by the use of certain matrices characterizing system properties: 
the controllability Gramian $P$, observability Gramian $Q$, required supply $\Pi$ and available storage $\Xi$:


\begin{itemize}
    \item Controllability Gramian $P$ is the unique solution to 
    \begin{equation}\label{eq:ctrb_lyap}
        0=AP + PA^\top + BB^\top,
    \end{equation}
    assuming $A$ to be a Hurwitz stable matrix. $P$ characterizes the least \emph{input} energy $L_P$ needed to steer the system state from $0$ to $x_0$ as $L_P(x_0) = x_0^\top P^{-1}x_0$. Eigenvectors corresponding to a large eigenvalue of $P$ thus require a small amount of energy to reach, i.e., they are greatly influenced by inputs and are in that sense `important'.
    \item Observability Gramian $Q$ is the unique solution to
    \begin{equation}\label{eq:obsv_lyap}
        0=A^\top Q + QA + C^\top C,
    \end{equation}
    assuming $A$ to be a Hurwitz stable matrix. $Q$ characterizes the observed \emph{output} energy $L_Q$ during the system's free evolution from $x_0$ to $0$, by $L_Q(x_0) = x_0^\top Qx_0$. Eigenvectors corresponding to a large eigenvalue of $Q$ are clearly observable in the output, i.e., they greatly influence outputs and are in that sense `important'.
    \item $\Pi$ is a (not necessarily unique) solution to 
    \begin{equation} \label{eq:dual_LMI}
        \left[\begin{array}{cc}
        A \Pi+\Pi A^\top & \Pi C^\top-B \\
        C \Pi-B^\top & -(D+D^\top)
        \end{array}\right] \preceq 0.
    \end{equation}
    As shown by \cite{Willems1972DissipativeTheory}, any solution $\Pi$ lies between two extremal solutions, i.e., $0\prec \Pi_{min} \preceq \Pi \preceq \Pi_{max}$. For balancing, we require the minimal solution $\Pi = \Pi_{min}$, which indicates the minimal amount of energy that must be added to the system to steer the system state from $0$ to $x_0$ by $L_\Pi(x_0) = x_0^\top \Pi^{-1}x_0$. Therefore, eigenvectors corresponding to a large eigenvalue of $\Pi$ are therefore `important'.
    \item $\Xi$ is a solution to \autoref{eq:pass_LMI}, which lies between two extremal solutions, i.e., $0\prec \Xi_{min} \preceq \Xi \preceq \Xi_{max}$ \citep{Willems1972DissipativeTheory}. For balancing, we require the minimal solution $\Xi = \Xi_{min}$, which indicates the mimimal amount of energy that can be recovered from the system over all state trajectories from $x_0$ to $0$ as $L_\Xi(x_0) = H(x_0)= x_0^\top\Xi x_0$. Eigenvectors corresponding to large eigenvalues of $\Xi$ are thus also `important'.
\end{itemize}

\begin{rem}
Both $\Pi$ and $\Xi$ indicate the passivity of the system $\sys$, such that the existence of a feasible $\Xi$ indicate the existence of $\Pi$ and vice-versa \citep{Gugercin2004AResults}.
\end{rem}
For the remainder of this paper, $\Pi$ and $\Xi$ refer to the minimal solutions $\Pi_{min}$ and $\Xi_{min}$, respectively. As $P$ and $\Pi$ relate to the steering of the state and $Q$ and $\Xi$ relate to the evolution of the state, $P$ and $\Pi$ are henceforth called \emph{input Gramians} and $Q$ and $\Xi$ \emph{output Gramians}.

Whether a system is in a balanced realization is directly related to the Gramians of the system.
\begin{defn}\label{defn:balsys_def}
    A minimal system $\sys$, as in \autoref{eq:default_SS}, is called ($X_i,X_o$)-balanced if $X_i$ is an input Gramian ($P$ or $\Pi$) and $X_o$ an output Gramian ($Q$ or $\Xi$) of system $\sys$, solving \autoref{eq:pass_LMI},\autoref{eq:ctrb_lyap}-\autoref{eq:dual_LMI}, where 
    \[
        X_i = X_o = \Gamma = \diag(\gamma_1,\dots,\gamma_n),
    \]
    where $\gamma_1\geq\dots\geq\gamma_n\geq 0$.
\end{defn}

A minimal system that is not balanced, can be transformed to a balanced realization as in Definition \ref{defn:balsys_def} using a similarity transformation. A similarity transformation is a transformation with a nonsingular matrix $T$, such that the transformed system becomes $\sysc = (TAT^{-1},TB,CT^{-1},D)$. 
\begin{thm}\label{th:balanced_Grams}
    \citep{Moore1981PrincipalReduction,Willems1972DissipativeRates} Consider a passive system $\sys$, as in \autoref{eq:default_SS}, and the realization $\sysc$, determined from a similarity transformation with nonsingular matrix $T$, as $\sysc = (\Ac,\Bc,\Cc,D)= (TAT^{-1},TB,CT^{-1},D)$. If the matrices $P$, $Q$, $\Pi$ and $\Xi$ satisfy the LMI's and Lyapunov equations of \autoref{eq:pass_LMI}, \autoref{eq:ctrb_lyap}-\autoref{eq:dual_LMI}
    for $\sys$, then $\Pc= T P T^\top$, $\Qc = T^{-\top} Q T^{-1}$, $\Pic = T \Pi T^\top$ and $\Xic = T^{-\top} \Xi T^{-1}$ satisfy the same equations for $\sysc$.
\end{thm}

The unique similarity transformation that results in the balanced realization of the system can be found by the algorithm of \autoref{tab:bal_alg}.

Since the Gramians are diagonal in a balanced system realisation, the `importance' of each state with respect to the specified Gramians' criteria is straightforward: the larger the diagonal entry, the higher the state's importance. The balanced system $\sysb$ is therefore partitioned as
\begin{equation}\label{eq:bal_part}
    \sysb=\left[\begin{array}{c|c}  \Ab & \Bb \\ \hline \Cb & \Db \end{array}\right]
    = \left[\begin{array}{cc|c}
        \Ab_{11} & \Ab_{12} & \Bb_{1} \\
        \Ab_{21} & \Ab_{22} & \Bb_{2} \\
        \hline \Cb_{1} & \Cb_{2} & D
    \end{array}\right],
\end{equation}
where the first partition corresponds to the states with the highest values of $\gamma_i$, as defined in Definition \ref{defn:balsys_def}. The reduced system $\sysr$ is obtained by selecting only the first partition, such that $\sysr\coloneqq (\Ar,\Br,\Cr,\Dr) = (\Ab_{11}, \Bb_1, \Cb_1, D)$. The reduced system $\sysr$ is also balanced, e.g., \cite{Gugercin2004AResults} and \cite{Unneland2007NewTruncation}. 

\begin{table}[]
    \caption{Balancing transformation algorithm,  \cite[Section~7.3]{Antoulas2005ApproximationSystems}}
    \label{tab:bal_alg}
    \centering
    \fontsize{6}{7}
    \begin{tabular}{l} \toprule
         (1) Pick input and output Gramians $X_i = X_i^\top \succ 0$, $X_o = X_o^\top \succ 0$. \\
         (2) Determine Cholesky factors as $X_i = R_i^\top R_i$,  $X_o = R_o^\top R_o$. \\ 
         (3) Perform the singular value decomposition $R_oR_i^\top = U\Gamma V$. \\
         (4) Set transformations as $T = \Gamma^{-1/2}U^\top R_o$, $T^{-1} = R_i^\top V\Gamma^{-1/2}$.\\
         (5) The balanced realizations is $\Ab = TAT^{-1}$, $\Bb = TB$, $\Cb = CT^{-1}$.\\
         (6) Gramians of the balanced realization, become $\tilde{X}_i = \tilde{X}_o = \Gamma$, \\ 
         \hspace{3.5mm} where $\tilde{X}_i = T X_i T^\top$ and $\tilde{X}_o = T^{-\top} X_o T^{-1}$.\\
         \bottomrule
    \end{tabular}
\end{table}

Standard terminology differentiates between three different balancing types, based on which Gramians are used, explained hereafter.
\vspace{-2mm}
\subsection[Lyapunov balancing]{($P$,$Q$)-balancing}\vspace{-2mm}
($P$,$Q$)-balancing, also known as Lyapunov balancing, is the original method as presented by \cite{Moore1981PrincipalReduction} and is still the most frequently used type of balancing. By using $P$ and $Q$, the $r<n$ states of the reduced-order model $\sysr$ all contribute significantly to the input-output behaviour, and the computation of $P$ and $Q$ is relatively cheap with respect to the computation of $\Pi$ and $\Xi$. The reduced system $\sysr$ is guaranteed to be stable if $\gamma_r > \gamma_{r+1}$ and an a priori $H_\infty$ error bound is available \citep{Antoulas2005ApproximationSystems}.
\vspace{-2mm}
\subsection[Positive real balancing]{($\Pi$,$\Xi$)-balancing} \label{ssec:PR_bal}\vspace{-2mm}
($\Pi$,$\Xi$)-balancing, also known as positive real (PR) balancing, selects states which contribute most to the internal energy of the system. Therefore, it is not as strongly related to the input-output behaviour of the system and is usually of lower accuracy than Lyapunov balanced truncation. However, positive real balancing guarantees the preservation of passivity and thus stability. An error bound is provided by \cite{Gugercin2004AResults}. 
\vspace{-2mm}
\subsection[Mixed-Gramian balancing]{($\Pi$,$Q$)- or ($P$,$\Xi$)-balancing} \label{ssec:MG_bal}\vspace{-2mm}
($\Pi$,$Q$)- and ($P$,$\Xi$)-balancing are both known as mixed Gramian (MG) balancing. Whether using ($\Pi$,$Q$) or ($P$,$\Xi$), the resulting balanced system is identical, as shown by \cite{Unneland2007NewTruncation}. MG balancing ensures the preservation of passivity through the one Gramian from PR balancing \citep{Unneland2007NewTruncation}. By using also a Gramian from Lyapunov balancing the accuracy of the frequency response is generally improved with respect to PR balancing. In addition, MG balancing is computationally less expensive than PR balancing as $P$ and $Q$ are generally cheaper to compute than $\Pi$ and $\Xi$. 


\section{Passivity-preserving, Interconnected Balancing}\label{sec:interc_systems}
In \autoref{sec:prob_def}, we set the aim as the reduction of the passive interconnected system $\sys_c$ by reduction of the subsystems, while ensuring passivity of $\sysr_c$. This preservation of passivity can be achieved straightforwardly by the reduction of each individual subsystem $\sys_j$ using either ($\Pi$,$\Xi$)-, ($\Pi$,$Q$)- or ($P$,$\Xi$)-balancing, as explained in \autoref{ssec:PR_bal} and \ref{ssec:MG_bal}. Even though this subsystem level reduction ensures passivity, the reduction of each $\sys_j$ does not take its environment (i.e., the other subsystems to which it connects) into account. As a consequence, the dynamics that is retained in each $\sysr_j$ might not contribute effectively to the accuracy of interconnected system $\sysr_c$.

To ensure that the reduced interconnected system $\sysr_c$ approximates $\sys_c$ accurately, while retaining the interconnection structure, \cite{Vandendorpe2008ModelSystems} present the Interconnected Systems Balanced Truncation (ISBT) method. In ISBT, the controllability and observability Gramians of the interconnected system are partitioned according to the state partition of \autoref{eq:state&IO_partitions}. For example, the controllability Gramian of the interconnected system $P_c$ is determined from
\begin{equation}\label{eq:cpld_ctrb_eqs}
    0=A_cP_c + P_cA_c^\top + B_cB_c^\top,\\
\end{equation}
where $P_c$ is partitioned as
\begin{equation}\label{eq:cpld_ctrb_part}
    P_c=\left[\begin{array}{ccc}
        P_{1,1} & \cdots & P_{1, k} \\
        \vdots & \ddots & \vdots \\
        P_{k, 1} & \ldots & P_{k, k}
    \end{array}\right],
\end{equation}
where $P_{i,j} \in \R^{n_i\times n_j}$. The same partition is applied to $Q_c$, the observability Gramian of $\sys_c$. ISBT then selects input Gramian $P_{j,j}$ and output Gramian $Q_{j,j}$ to transform $\sys_j$ using the balancing algorithm of \autoref{tab:bal_alg}. The resulting realization $\sysb_j$ is called \emph{subsystem balanced}. Note that the subsystem itself is not ($P_j$,$Q_j$)-balanced, i.e., the Gramians $P_j$ and $Q_j$ of $\sys_j$ are not equal and diagonal. The subsystem-balanced subsystems $\sysb_j$ are truncated by selection of the first partition, as in \autoref{eq:bal_part}. The main drawback of ISBT is that stability and passivity of $\sysr_c$ are generally not preserved and there is no error bound available. 

As the main contribution of this paper, we propose a new method which combines the advantages of previous research on interconnected systems and passivity. Specifically, a combination of ISBT and ($P$,$\Xi$)-balancing is proposed to accurately reduce interconnected systems, guaranteeing passivity (and therefore stability) of both subsystems $\sysr_j$ and interconnected system $\sysr_c$. To this end, we solve and partition $P_c$ as in \autoref{eq:cpld_ctrb_eqs} and \autoref{eq:cpld_ctrb_part}, and we determine the available storage $\Xi_j$ of subsystem $\sys_j$ from
\begin{equation}\label{eq:subsys_stor}
       \left[\begin{array}{cc}
            A_j^{\top} \Xi_j+\Xi_j A_j & \Xi_j B_j-C_j^\top \\
            B_j^\top \Xi_j-C_j & -(D_j+D_j^\top)
        \end{array}\right] \preceq 0.
\end{equation}
$P_{j,j}$ and $\Xi_j$ are then used to transform $\sys_j$ to $\sysb_j$, using the algorithm of \autoref{tab:bal_alg}. By partitioning of $\sysb_j$ as in \autoref{eq:bal_part}, and retention of only the $(\Ab_{11}, \Bb_1, \Cb_1, D)$-partition, the reduced-order subsystem $\sysr_j$ is obtained. The reduced-order subsystems can then be interconnected using $S$ and $\mathcal{B}$ to obtain $\sysr_c$. We call this new reduction scheme \emph{Passive Interconnected Balanced Truncation} (PIBT), which is summarized by \autoref{alg:PIBT}.
\begin{alg}{PIBT}\label{alg:PIBT}\\
Consider $k$ passive, minimal subystems $\sys_j$, coupling matrices $S$ and $\mathcal{B}$ and the asymptotically stable system $\sys_c$.
    \begin{enumerate}
        \item Calculate the global controllability Gramian $P_c$ from \autoref{eq:cpld_ctrb_eqs} and partition it according to the state, as in \autoref{eq:cpld_ctrb_part}.
        \item Calculate, for each subsystem $\sys_j, \ j \in \{1,\dots,k\}$, its available storage $\Xi_j$ using \autoref{eq:subsys_stor}.
        \item For each subsystem $\sys_j$, employ the algorithm of \autoref{tab:bal_alg} with $X_i = P_{j,j}$ and $X_o = \Xi_j$ to obtain the balanced $\sysb_j$.
        \item Truncate the balanced system $\sysb_j$ to $\sysr_j$ by selecting only the $(\Ab_{11}, \Bb_1, \Cb_1, D)$-partition of the partitioned, balanced system, as in \autoref{eq:bal_part}.
        \item Interconnect all the reduced subsystems $\sysr_j$ with $S$ and $\mathcal{B}$ according to \autoref{eq:cpld_sys} and \autoref{eq:cpld_sys_matrices}, to find $\sysr_c$.
    \end{enumerate}
\end{alg}

The PIBT algorithm of \autoref{alg:PIBT} guarantees the passivity of both the reduced subsystems $\sysr_j$ and reduced interconnected system $\sysr_c$, as shown in the following theorem.
\begin{thm}
    Consider a passive, minimal interconnected system $\sys_c$ as in \autoref{eq:cpld_sys}, consisting of $k$ passive, minimal subsystems $\sys_1,\dots,\sys_k$, a positive semi-definite interconnection matrix $S$ and external input matrix $\mathcal{B}$. If $\sys_c$ is reduced to $\sysr_c$ using PIBT of \autoref{alg:PIBT}, $\sysr_j, \  j\in\{1,\dots,k\}$ and $\sysr_c$ are passive.
\end{thm}
\begin{pf}
    By the balancing algorithm of \autoref{tab:bal_alg}, $\Xi_j$ is transformed to $\Xib_j = \Gamma$, i.e., $\Xib_j$ is diagonal and sorted. Therefore, for the balanced subsystem the LMI in \autoref{eq:pass_LMI} can be partitioned as
    \begin{equation}
    \label{eq:part_bal_stor_ric}
        \left[\begin{array}{cc}
            L_{11} &  L_{12}\\
            L_{21}& -(D_j+D_j^\top)
        \end{array}\right]\preceq 0,
    \end{equation}
    where
    \begin{equation}\fontsize{9}{11} \label{eq:part_bal_stor_mats}
    \begin{aligned}
        L_{11} &= \begin{bmatrix} \Ab_{j,11} & \Ab_{j,12}\\ \Ab_{j,21}& \Ab_{j,22} \end{bmatrix} ^\top 
            \begin{bmatrix} \Gamma_1 & O\\ O & \Gamma_2 \end{bmatrix} 
            + \begin{bmatrix} \Gamma_1 & O\\ O & \Gamma_2 \end{bmatrix} 
            \begin{bmatrix} \Ab_{j,11} & \Ab_{j,12}\\ \Ab_{j,21}& \Ab_{j,22} \end{bmatrix}\\
        L_{12} &= \begin{bmatrix} \Gamma_1 & O\\ O & \Gamma_2 \end{bmatrix} 
            \begin{bmatrix} \Bb_{j,1}\\ \Bb_{j,2} \end{bmatrix} -
            \begin{bmatrix} \Cb_{j,1}^\top\\ \Cb_{j,2}^\top \end{bmatrix}
    \end{aligned}
    \end{equation}
    At the truncation step, $\Xib_j = \Gamma$ is truncated to $\Xir_j = \Gamma_1$. The LMI for the available storage of the reduced system $\sysr_j$ corresponds to \autoref{eq:part_bal_stor_ric} with only the top-left and top partitions of $L_{11}$ and $L_{12}$ in \autoref{eq:part_bal_stor_mats}, respectively. The truncated available storage $\Xir_{av,j}$ is a valid solution to this LMI, such that $\sysr_j$ retains the passivity properties of $\sys_j$. $\sysr_c$ is therefore an interconnection of the passive subsystems $\sysr_j$ and interconnection matrices $S$ and $\mathcal{B}$ and is thus also passive according to \autoref{th:interconnection_passivity}.   
    \hfill$\square$
\end{pf}

\begin{rem}
    Note that in \autoref{alg:PIBT}, the balancing algorithm uses $P_c$ and $\Xi_j$. A valid alternative would be to use $\Pi_j$ and $Q_c$, which also guarantees passivity of the reduced systems, but generally results in a different reduced order model.
    Only if all subsystems are fully symmetric, i.e. $C_j = B_j^\top,$ for all $j \in \{1,\dots,k\}$, then $P_c = Q_c$ and $\Xi_j = \Pi_j$, such that the choice between $P_c$ and $\Xi_j$ or $\Pi_j$ and $Q_c$ is equivalent. Numerical tests on non-symmetric systems have demonstrated comparable accuracy for either $P_j$ and $\Xi_j$ or $\Pi_j$ and $Q_c$, but further study is required to provide guidelines on which to use.
\end{rem}



\section{Numerical example}\label{sec:num_example}
\begin{figure}
    \includegraphics[width = 0.9\linewidth]{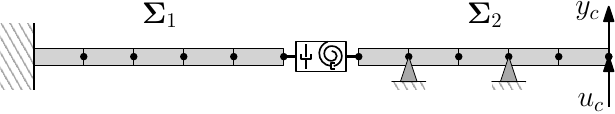}
    \vspace{-2mm}
    \caption{Schematic drawing of two interconnected Euler beams.}
    \label{fig:num_example}
\end{figure}

As an illustrative example, we compare three reduction methods by their application to two interconnected beam models, $\sys_1$ and $\sys_2$, as schematically shown in \autoref{fig:num_example}. Both models represent 1 m long, steel beams (Young's modulus is $2\times 10^{11}$ Pa, mass density is $8\times 10^{3}$ kg/m$^3$) with a square cross-section area of $1\times 10^{-4}$ m$^2$, consisting of 5 identical 2D Euler beam elements (subsystem order $n_j = 20$, for $j = 1,2$). The left beam, modeled by $\sys_1$, is clamped at its left end, i.e., it is a cantilever beam. The right beam, modeled by $\sys_2$, has its second and fourth transversal translational degrees of freedom fixed, as shown in \autoref{fig:num_example}. Rayleigh damping is added such that the modal damping parameters are given as $\xi_k = 0.5(\omega_{Ok}^{-1} + 5\times 10^{-6} \omega_{Ok})$, with $\omega_{Ok}$ the undamped angular eigenfrequencies. By using solely collocated force inputs and velocity outputs, both beam models are passive.

The two beams are interconnected by a transversal translational damper and a rotational damper with damping constants of $50$ Ns/m and $3$ Nms/rad, respectively. The interconnected system $\sys_c$ has one external input force $u_c$ and output velocity $y_c$ at the right free end of the right beam and is therefore also passive.

The subsystems models $\sys_j$ are reduced to $\sysr_j$, both of order $12$, such that $\sysr_c$ is of $24$th order. This reduction is performed using three reduction methods:
\begin{itemize}
    \item Mixed-Gramian balanced truncation (MGBT) of \cite{Unneland2007NewTruncation} to reduce the individual subsystems using ($P,\Xi$)-balancing. In other words, both reductions $\sys_1\rightarrow\sysr_1$ and $\sys_2\rightarrow\sysr_2$ are performed individually before coupling to consitute $\sysr_c$.
    \item Interconnected systems balanced truncation (ISBT) of \cite{Vandendorpe2008ModelSystems} to reduce the subsystems based on the Gramians of the interconnected system $\sys_c$.
    \item Passive interconnected balanced truncation (PIBT) to reduce the subsystems as presented in \autoref{alg:PIBT}.
\end{itemize}

\begin{figure}
    \centering
    \includegraphics[width = 0.9\linewidth]{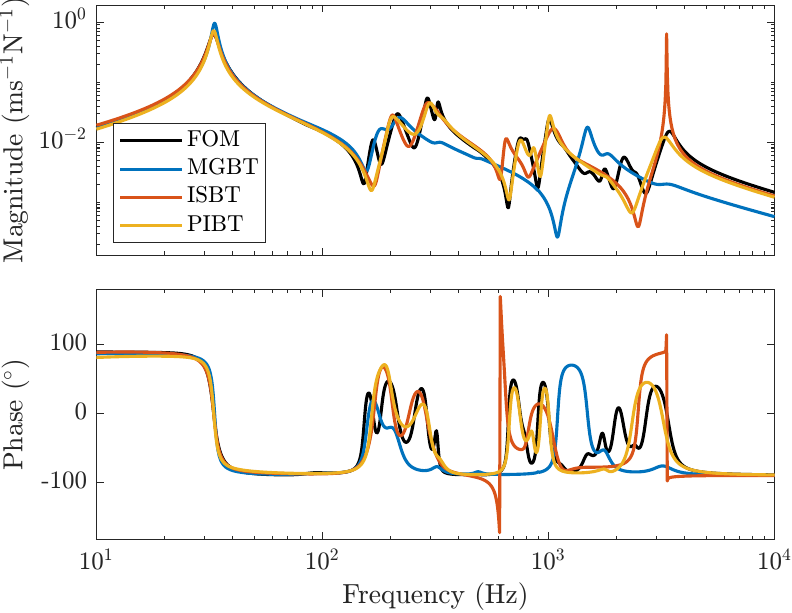}
    \centering \vspace{-2mm}
    \caption{FRF of the full-order model (FOM) $\sys_c$ and the three reduced order models.}
    \label{fig:FRF}
\end{figure}

The three reduced-order models $\sysr_c$ (ROMs) are compared to the full-order model $\sys_c$ (FOM) by means of their frequency response functions (FRFs), norms on the corresponding error, and their step response. The systems $\sys_c$ and $\sysr_c$ can alternatively be characterized by their transfer functions $G_c(s) \coloneqq C_c(sI-A_c)^{-1}B_c+D_c$ and $\Gr_c(s) \coloneqq \Cr_c(sI-\Ar_c)^{-1}\Br_c+\Dr_c$, respectively. The error can then be defined as $E(s) \coloneqq G_c(s) - \Gr_c(s)$. To characterize this error, we evaluate its H$_2-$ and L$_\infty$-norm \citep{Zhou1998EssentialsControl} as
\begin{equation}
    \|E\|_2^2 = \frac{1}{2\pi} \int_{-\infty}^{\infty} \text{trace}[E^*(j\omega)E(j\omega)] \text{d}\omega,
\end{equation}
\begin{equation}
    \|E\|_\infty \coloneqq \text{ess }\underset{\omega \in \R}{\text{sup}}\ \bar{\sigma}[E(j\omega)].
\end{equation}
Note that $\|E\|_2$ is only defined if $E$ is stable and is defined to be infinity otherwise. $\|E\|_\infty$ is only defined if $E$ has no purely imaginary poles and otherwise as infinity. Therefore, the L$_\infty$-norm, as opposed to the H$_\infty$-norm, is also defined for unstable systems.

The first comparison of the ROMs, by means of the FRFs shown in \autoref{fig:FRF}, indicates that PIBT approximates the FOM most accurately. The FRF of the MGBT-ROM
follows the FRF of the FOM the least accurately, as MGBT operates
on the individual subsystems $\sys_j$ instead of $\sys_c$, like with ISBT and PIBT. Still, the FRF of the ISBT-ROM shows arguably larger deviation than the PIBT-ROM, e.g., near 700 Hz and 3300 Hz.


A second comparison based on norms on the errors shown in \autoref{tab:err_nrms}, indicates PIBT performs best, showing both the smallest L$_\infty$- and the smallest H$_2$-error norm. Additionally, in contrast to the PIBT-ROM, the ISBT-ROM shows an infinite H$_2$-norm of its error, indicating its instability. A further check shows that, whereas the ISBT-ROM is neither passive nor stable, the MGBT- and PIBT-ROMs retain both stability and passivity, which is in line with the theoretical guarantee of \autoref{th:interconnection_passivity}.

\begin{figure}
    \centering
    \includegraphics[width = 0.9\linewidth]{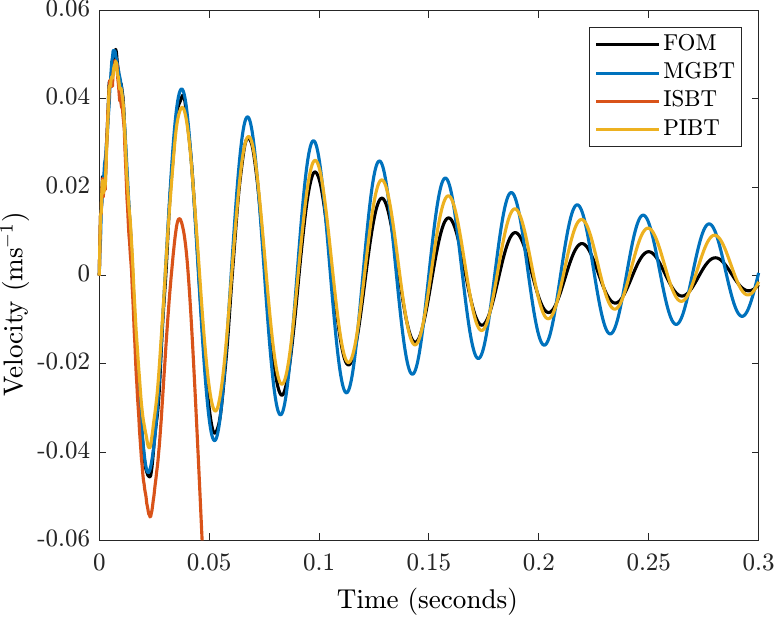}\vspace{-2mm}
    \caption{Step response of the full order model (FOM) $\sys_c$ and the three reductions.}
    \label{fig:step}
\end{figure}

The final comparison, based on the step responses shown in \autoref{fig:step}, confirms the superiority of PIBT as observed in above two comparisons. Whereas the PIBT-ROM's step response matches the FOM's step response more accurately than the MGBT-ROM does, the step response of the ISBT-ROM diverges within the first 0.05 s, again indicating instability.

\begin{table}
    \centering
    \caption{H$_2$-norm and L$_\infty$-norm of $E(s) \coloneqq G_c(s) - \Gr_c(s)$ for the three reduction methods.}
    \vspace{-2mm}
    \begin{tabular}{lccc}\toprule
        & MGBT & ISBT & PIBT \\ \midrule
        H$_2$-norm [m$^2$/(N$^2$s$^3$)]& $1.13$ & $\infty$ & $0.463$ \\
        L$_\infty$-norm [m/(Ns)] & $0.381$ & $1.48$ & $0.0950$ \\ \bottomrule
    \end{tabular}
    \label{tab:err_nrms}
\vspace{2mm}
\end{table}
Overall, the newly presented method of PIBT demonstrates superior performance with respect to both MGBT and ISBT in this numerical example. Where MGBT compromises on accuracy and ISBT loses stability, PIBT shows it can combine both a high accuracy and a passivity/stability guarantee.

\vspace{-1mm}
\section{Conclusions and Recommendations}\label{sec:conclusion} 
\vspace{-1mm}
In order to preserve both passivity and internal structure, interconnected systems are usually reduced by the individual reduction of its subsystems, which does not guarantee the accuracy of the reduced interconnected system. We have presented Passive Interconnected Balanced Truncation (PIBT) as an alternative approach for the reduction of an interconnected system in \autoref{alg:PIBT}. This reduction method works on the interconnected system's model to allow higher accuracy, while preserving the interconnection structure and both the passivity of subsystem models and the interconnected system's model. In the presented numerical example, PIBT shows a significantly improved approximation of the interconnected system response compared to other reduction methods. A practical limitation of PIBT is the need for LMI solutions, which is computationally less attractive for large subsystem models. For future research, we will work on a more computationally efficient approach, to solve these LMI's. In addition, we will generalize PIBT to be more widely applicable and study the feasibility of defining error bounds.
\vspace{-0mm}

\begin{ack}
The authors would like to thank the company ASML for its financial support, and specifically Dr. Victor Dolk and Thijs Verhees, MSc, for valuable discussions.
\end{ack}        
 
\bibliography{references}

\end{document}